\documentclass[aps,prb,groupedaddress,twocolumn, amsmath,amssymb,showpacs,
eqsecnum,floatfix]{revtex4}

\usepackage{graphicx}
\begin{document}


  
\title{Phase diagram of randomly pinned vortex matter 
in layered superconductors: dependence on the details of
the point pinning} 

\author{Chandan Dasgupta}
\email{cdgupta@physics.iisc.ernet.in}
\altaffiliation{Also at Condensed Matter Theory Unit, Jawaharlal Nehru Centre
for Advanced Scientific Research, Bangalore 560064, India}
\affiliation{Centre for Condensed Matter Theory, Department of Physics, 
Indian Institute  of Science, Bangalore 560012, India}
\author{Oriol T. Valls}
\email{otvalls@umn.edu}
\altaffiliation{Also at Minnesota Supercomputer Institute, University of Minnesota,
Minneapolis, Minnesota 55455}
\affiliation{School of Physics and Astronomy,
University of Minnesota, Minneapolis, Minnesota 55455}

\date{\today}

\begin{abstract}
We study the thermodynamic and structural properties of the superconducting 
vortex system in high temperature layered superconductors, with
magnetic field normal to the layers, in the presence of a small
concentration  of strong random point pinning defects 
via numerical minimization of a model
free energy functional in terms of the time-averaged local density
of pancake vortices. Working at constant magnetic induction and point
pinning center concentration, we find that the equilibrium phase at low  
temperature ($T$) and small pinning strength
($s$) is a topologically ordered Bragg glass. As $T$ or $s$ is increased,
the Bragg glass undergoes 
a first order transition to a disordered phase
which we characterize as a ``vortex slush'' 
with polycrystalline structure within
the layers and interlayer correlations extending to about twenty layers.
This is in contrast with the pinned vortex liquid 
phase into which the Bragg glass
was found to melt, using the same methods, in the case of a large concentration
of weak pinning centers: that phase was amorphous with very little interlayer
correlation. The value of the second moment of the random 
pinning potential at which the  Bragg glass melts for a fixed 
temperature is very  different in the two systems. 
These results imply that the
effects of random point pinning can not be described only 
in terms of the second moment
of the pinning potential, and that some of the
unresolved contradictions in the
literature concerning the nature of the low
$T$ and high $s$  phase in this system are likely to arise from differences
in the nature of the pinning in different samples, or 
from assumptions made about
the pinning potential.

\end{abstract}

\pacs{74.25.Qt, 74.72.Hs, 74.25.Ha, 74.78.Bz}

\maketitle

\section{Introduction}
\label{intro}

The effects of random point pining on the properties of vortex matter in
high-temperature superconductors have been extensively investigated 
\cite{review} in a large number of experimental, theoretical, and numerical
studies. However, many questions about the phase diagram of this system
remain controversial. The existence of a topologically ordered Bragg glass
(BrG) phase~\cite{natt,giamarchi} with long-range bond-orientational order 
and power-law decay of translational correlations
at low temperatures and low fields in systems with weak pinning has
been experimentally established~\cite{klein}. The BrG phase melts into a
disordered vortex liquid (VL) as the temperature is 
increased. It also 
becomes unstable as the pinning strength is 
increased (or equivalently, the magnetic field is increased 
for a fixed amount of pinning) with the temperature held fixed at a low
value. There is no consensus at present about the nature of the phase 
into which the BrG transforms under these conditions. The theoretical
proposal~\cite{mpa,ffh} of the existence of a vortex glass (VG) with
long-range coherence of the superconducting phase and divergent barriers
for the motion of vortex lines in the limit of zero current has received
support from many experiments~\cite{vglexp,zeldov1,zeldov2}. However, 
other experimental studies\cite{lobb} have questioned the
the existence of a VG phase thermodynamically distinct from the
high-temperature VL. Different numerical studies~\cite{bokil,olsson,lidmar,nh}
have also reached different conclusions about the existence of the VG phase.

The possible existence of other disordered phases, distinct from the 
VL and VG phases mentioned above, has also been suggested in several studies.
One of these phases is the ``vortex slush'' (VS) -- a disordered, liquid-like
phase with no superconducting phase coherence that is distinguished from the
VL in that the translational correlation length in this phase is substantially
larger than that in the VL. The existence of this phase in superconductors
with random point pinning has been predicted 
theoretically~\cite{slush_th1,slush_th2}  
and several experimental studies~\cite{slush1,slush2,slush3} have
presented evidence for the occurrence of a first-order transition between
VL and VS phases (followed by a continuous transition from the VS to the VG
phase) as the temperature is decreased at fixed magnetic field.
A numerical study~\cite{nh} has also provided evidence for the existence of
the VS phase, although the validity of this
evidence for macroscopic systems has been questioned 
in later work~\cite{teitel}.

Another glassy phase proposed to exist in both high-temperature
and conventional superconductors with random point pinning is the 
``multidomain glass'' (MG)~\cite{banerjee,gautam} in which
the vortices are supposed to form 
crystalline domains with typical size  substantially larger than the
translational correlation length in the VL phase. In this MG phase, the vortex
positions do not exhibit any long-range translational or orientational
order and there is no superconducting phase coherence.
It has been argued
~\cite{banerjee,gautam} that a variety of interesting ``glassy'' behavior
observed~\cite{banerjee} near the melting transition 
of the BrG phase may be explained by
assuming that a small sliver of MG phase exists between the BrG and VL
phases, so that the BrG first transforms to the MG phase as the temperature 
is increased,
and then the MG melts into the VL at a slightly higher temperature. 
Results of recent muon-spin-rotation experiments~\cite{divakar,menon2}
have been interpreted as evidence for the occurrence of vortex structures
similar to that expected in the MG phase in the disordered glassy phase
of a high-temperature superconductor. 

There is an
additional controversy about the structure of the disordered vortex
state at low temperatures and strong pinning (or 
high magnetic field). Early decoration
experiments~\cite{deco1} on layered high-temperature 
superconductors showed evidence
for an amorphous arrangement of vortices in the disordered phase.
However, more recent experiments~\cite{deco2}
on a low-temperature superconductor, ${\rm NbSe_2}$, 
have shown the occurrence of polycrystalline
disordered structures. A similar disagreement exists between the results of
different simulations: while a simulation~\cite{sim1} of the structure of a
two-dimensional vortex system with random point pinning shows a
polycrystalline arrangement of the vortices, another numerical
study of a similar
system~\cite{rodriguez} has found amorphous structures without
well-defined crystalline regions separated by grain boundaries. The reason
for this disagreement 
remains to be understood.

The question we wish to address 
in this work is whether the differences
(summarized above) among the results of different studies of the properties
of vortex matter with random point pinning arise, at least in part, from
differences in the details of the pinning potential. 
It is often quite difficult
to obtain reliable information about the microscopic pinning centers present
in an experimentally studied sample. Therefore, it is possible that 
samples of the same superconductor studied in different experiments have 
different pinning potentials. This may account for
some of the differences in the experimental results. Similarly, the pinning
potentials used in different simulations are often quite different from
one another. It is, therefore, important to determine the extent to which
the details of the pinning potential are relevant in determining the
properties of the system.
Here we tackle  this issue
by  studying  the thermodynamic 
and structural properties
of the mixed phase of a highly anisotropic layered high-temperature 
superconductor in the presence of a {\it small} 
concentration of randomly located
{\it strong} point pinning centers. The 
pinning centers on different layers are
assumed to be completely uncorrelated.  
The primary motivation for this study is to examine the dependence of the
phase diagram of this system and the structure of the disordered phase on 
the details of the random pinning potential by comparing the results
obtained with previous results\cite{dv06} obtained from using the same methods
for the same system, but with a much larger concentration of weaker point
defects. 

In theoretical studies, the random 
pinning potential $v({\bf r})$ is usually assumed
to be a random Gaussian variable with zero mean, 
whose statistics is completely specified
by the second moment of its distribution, $\langle v({\bf r}) v({\bf r}^\prime)
\rangle = K^2 f(|{\bf r} - {\bf r}^\prime|)$, where $f(x)$ is a short-ranged
function (its range is usually assumed to be 
of the order of the superconducting coherence 
length $\xi$) normalized to unity at $x=0$, and the parameter $K$ measures
the strength of pinning. For pinning due to a collection
of randomly distributed point pinning centers, $K$ is proportional to the
product of the depth $s$ of the pinning potential of an 
individual pinning center and the
square root of the concentration $c$ of pinning centers. Thus, two systems,
one with small $s$ and large $c$ (i.e. a large concentration of weak pinning
centers) and a second one with large $s$, but small $c$ (i.e. a small
concentration of strong pinning centers) may have very similar values of
the parameter $K$, so that an analytical treatment based on the assumption
of Gaussian randomness would predict the same behavior for the two systems.
Physically, however, it is not al all obvious that 
the properties of these two systems would be
very similar. The main objective of our study is to examine this question. 

Our study proceeds by numerical minimization of a discretized version of 
the Ramakrishnan-Yussouff (RY)~\cite{ry} 
free-energy functional for the system of pancake vortices in a highly 
anisotropic layered high-temperature superconductor, in 
a magnetic field normal to the layers. Different phases of the system
correspond to different local minima of the free energy in this mean-field
description, and a crossing of the free energies of two different 
minima represents a first-order phase transition. Information about the
structure of different phases is obtained from various correlation functions
of the density distributions at the corresponding free-energy minima. 
We have previously used this method to study the thermodynamic 
and structural properties of this system in the presence of a small
concentration of strong columnar pinning centers perpendicular to the 
layers~\cite{us1,us2,prbd}, and also a large concentration of randomly placed
weak point pinning centers~\cite{dv06}. We found the phase diagrams in the two 
cases to be quite different. In the system with a small 
concentration of strong columnar pins, the low-temperature BrG phase 
exhibited a two-step melting, via an intermediate polycrystalline
Bose glass phase, to the VL. This is
qualitatively similar to the prediction~\cite{banerjee,gautam} of two-step
melting (via an intermediate MG phase) of the BrG phase in systems with
random point pinning.  
In contrast, we found a single first-order melting of the BrG
phase in the system with a large concentration of weak random point
pinning centers. The structure of the disordered phase 
was amorphous in this case. 

The pinning potentials in these
two cases differ in two important ways. First, the 
pinning potential on different layers are perfectly correlated for columnar
pins and completely uncorrelated for point pinning. Second, the system with
columnar pins has a small concentration of strong pining centers, whereas the
one with point pinning has a large concentration of weak pinning centers.
It is not clear which one of these two differences is the primary reason for
the observed differences in the properties of these two systems. More
important is the possibility that, in the
case of point pinning alone, differences in the concentration and strength
of the pinning centers can account for the apparently conflicting behaviors
discussed above.
To shed  light on these issues, we  consider
here a system with a
small concentration of strong pinning centers whose positions
on different layers are completely uncorrelated. The values of the areal
concentration $c$ of pinning centers on each layer and the strength $s$ 
of the pinning potential of each pinning center are chosen so as to include
the region where the values of the parameter $K \propto \sqrt{s^2c}$  are
similar to those for the system
with large $c$ and small $s$ studied earlier~\cite{dv06}. Thus, a comparison
of the results obtained here with those of 
Ref.~\onlinecite{dv06} can provide useful information about the
dependence of the properties of disordered vortex matter on the details of the
random pinning potential. 
The model of pinning considered here 
may be appropriate for films of high-temperature superconductors 
in which meandering lines
of dislocations~\cite{film1,film2} or artificially introduced material defects
~\cite{film3} act as strong columnar 
pinning centers that are not perfectly
correlated across layers. 

We consider a fixed value of the magnetic induction $B$ = 2kG
and study the phase behavior for different values of the strength $s$ of the
pinning potential, keeping the low pin concentration $c$ fixed
(a discussion of
how the behavior is expected to depend on the value of $B$ may be found in
Ref. \onlinecite{dv06}).
For relatively small values of $s$, we find a topologically ordered BrG phase
at low temperatures, which melts into a VL via a first-order transition 
as $T$ is increased. As  $s$ is increased, the 
temperature at which this melting transition occurs decreases, and the 
transition line in the $(T-s)$ plane tends to become parallel to the
$T$-axis for large $s$, indicating that the BrG phase does not exist if the 
pin strength exceeds a critical value. We find that
the disordered phase into which the
BrG transforms as $s$ is increased at low $T$ is glass-like:
the vortices are strongly localized at points that do not form a
structure with long-range translational or orientational correlation. However,
this phase transforms continuously into the high-temperature VL as $T$ is
increased at constant $s$: we do not find any evidence for a phase transition
between this glassy state and the VL.

Although the general features of the phase diagram are 
qualitatively similar to those
found in our earlier study,~\cite{dv06}
several other features found here
are substantially different. 
The structure of the glassy state at large $s$ and low $T$ 
is quite different from that of the disordered phase found
in Ref.~\onlinecite{dv06}. In the glassy phase of the present system, 
the vortices on each layer form a polycrystalline structure with crystalline
domains separated by grain boundaries, in contrast to the amorphous structure
found in Ref.~\onlinecite{dv06}. Also, the vortex positions on different layers
are now significantly correlated, with a correlation length
of the order of 20 layer spacings. This is very different from the 
previously found\cite{dv06} glassy
state, which does not exhibit any interlayer
correlation in the vortex positions. The glassy state found here has the
characteristics of the VS and MG phases mentioned above. For this reason, we
classify this state as a VS, whereas the glassy state found in 
Ref.~\onlinecite{dv06} was identified as a pinned vortex liquid. 
Another 
difference in the properties of these two systems lies in the location of a 
crossover line that separates the region in the $(T-s)$ plane in which the
VS is ``glassy'' 
in the sense that the peak vortex densities  are
high from that in which these
densities are more liquid-like. This crossover line is
obtained using a criterion~\cite{prbd} based on percolation of liquid-like
regions. The location of this crossover line in relation to the 
first order line in 
the $(T-s)$ plane bounding the  BrG phase
is quite different in the two systems. We also find 
that the rms value of the random pinning potential at which the BrG melts
in the present system 
is about an order of magnitude smaller than that at which the BrG to pinned
vortex liquid transition occurs in the system of Ref.~\onlinecite{dv06} at
the same temperature. These
differences illustrate the importance of the details of the pinning potential
in determining the properties of disordered vortex matter, and suggest that
the predictions of analytic calculations in which the effects of random 
pinning are assumed to be described completely by the second moment of the
distribution of the pinning potential may not be quantitatively correct.

The rest of this paper is organized as follows. The model considered here
and the method of calculation are described in section~\ref{methods}. In
section~\ref{results}, we describe in detail the results of our calculations
and compare these results with those obtained in Ref.\onlinecite{dv06}.
Section~\ref{summary} contains a summary of our main results and a few
concluding remarks.
 
\section{Model and Methods}
\label{methods}

We consider the system of pancake vortices in a highly anisotropic layered
superconductor with vanishingly small Josephson coupling between layers  
in a magnetic field perpendicular to the layers. In this system, the pancake 
vortices on different layers are coupled only 
via their electromagnetic interaction.
Our starting point is the expression for
the free energy of the vortex system as a
functional of the time-averaged  areal
density of pancake vortices $\rho_n({\bf r})$. Here ${\bf r}$ is a two
dimensional vector and the discrete index $n$ numbers the layers.
We write
\begin{equation}
F[\rho]=F_{RY}[\rho]+F_p[\rho]
\label{fe}
\end{equation}
where the first term
in the right-hand side  is the free energy 
in the absence
of pinning, while the second includes the
pinning effects. For the first term we use
the RY  form~\cite{ry}:
\begin{widetext}
\begin{equation}
\beta F_{RY}[\rho]= 
\sum_{n}\int{d^2 {\bf r}\{\rho_n({\bf r})
\ln (\rho_n({\bf r})/\rho_0)-\delta\rho_n({\bf r})\} } 
(1/2)\sum_m \sum_n \int{d^2 {\bf r} \int {d^2{\bf r}^\prime \,
C_{mn}({|\bf r}-{\bf r^\prime|}) \delta \rho_m ({\bf r}) \delta
\rho_n({\bf r}^\prime)}},
\label{ryfe}
\end{equation}
\end{widetext}
where $\beta$ is
the inverse temperature and
$\delta \rho_n ({\bf r})\equiv \rho_n({\bf r})-\rho_0$  the
deviation of  ${\rho_n(\bf r})$ from its average value $\rho_0$,
the density of the uniform liquid ($\rho_0 \equiv B/\Phi_0$ 
where $B$ is the 
magnetic induction and $\Phi_0$  the superconducting flux quantum). We
have taken our zero of the free energy at its uniform liquid value.
$C_{mn}(r)$ denotes the direct pair correlation function
of the uniform vortex liquid
at density $\rho_0$. This static
correlation function is assumed known and
contains  the required information about the interactions.
We have taken for $C_{mn}(r)$ the expression obtained from the
hypernetted chain approximation in
Ref.~\onlinecite{menon1}. We use parameter values appropriate to
BSCCO, taking the same numerical values as in Ref.~\onlinecite{dv06}.

For the second (pinning) term in Eq.~(\ref{fe}) we write:
\begin{equation}
F_p[\rho]= \sum_n \int{d {\bf r} V^p_n({\bf r}) [\rho_n({\bf r})-\rho_0] },
\end{equation}
where $V^p_n({\bf r})$ is the pinning potential
at point ${\bf r}$ on layer $n$.
The pinning potential is assumed to be produced by
random atomic scale point defects.
We characterize the  concentration  of these point defects
by the fraction $c$ of the
cells in the underlying BSCCO {\it crystal lattice} (which we take to be
orthorhombic with in-plane lattice spacing $d_0= 4\AA$ and 
interlayer distance $d=15\AA$) that contain defects.

To study the properties 
of the vortex system, we discretize the free energy
by introducing variables $\rho_{n,i}$ in a triangular computational lattice
of size $N^2\times N_L$ with periodic boundary conditions, 
where $N_L$ is the number of layers. Here 
$\rho_{n,i}=A_0\rho_n({\bf r}_i)$ where $A_0$ is the area of the in-plane unit
cell of the computational lattice and ${\bf r}_i$ denotes the location of the
$i$th computational lattice point in the $n$th layer. 
We then minimize the free energy functional
with respect  to the variables  $\rho_{n,i}$ using previously described
procedures.\cite{cdo,prbv}
All of the results presented here are for computational
lattices with $N=256$ and $N_L=128$.
We work at constant field $B=0.2T$.
We take the computational 
lattice constant to be $h=a/16$, where $a$ denotes the
equilibrium value\cite{prbv}
of the lattice constant of the vortex lattice of the pure system
for $B=0.2T$ near its melting transition.
With these choices, the number of
pancake vortices per layer is $N_v=256$.
All lengths are measured in units of $a_0$
defined by $\pi\rho_0 a_0^2=1$.

To determine the pinning potential
$V_{n,i}$ associated with each computational cell, we generate in each
layer a random set of $N_d$  points, where $N_d$ is the number of defects
per layer, $N_d=cN^2A_0/d_0^2$. We then have $V_{n,i}=s(m_{n,i}-N_d/N^2)$
where $s$ is the pinning potential strength, which we will give in
degrees $K$, and $m_{n,i}$  the number of defect  
points associated with the $(n,i)$
computational site. We consider here  relatively
large values of $s/T$ and  small values of $c$ (see below). In that case
the most frequent value of the integer
$m_{n,i}$  is zero and it rarely exceeds unity. With these conventions
the average of $V_{n,i}$ is zero while its fluctuations are
$\langle V_{n,i}V_{n^\prime,j} \rangle =
268.3 s^2 c \delta_{n,n^\prime}\delta_{i,j}$. 
In most of our calculations, the  concentration $c$ of
pinning defects was fixed at a value that corresponds to having 
$N_d=24$ pins per layer, but we have also checked
that results for 32 and 64 pins per layer are qualitatively the same
and quantitatively very similar.
The pin concentration of 24 per layer corresponds, in physical units,
to having $1.36\times 10^{-4}$
pinning defects for every hundred unit cells
in the underlying {\it crystal lattice}. This is in contrast
to the value used in Ref.~\onlinecite{dv06} which was one defect per hundred
unit cells of the crystal lattice. 
Thus the value of $c$ used here is about four orders of magnitude smaller.
In contrast we use values of the strength $s$ one or two orders of magnitude
larger, so that the product $s^2c$ that determines the second moment 
of the random pinning potential is comparable in the two cases.

\section{Results}
\label{results}

In this  section we discuss the results obtained using the
methods described in the preceding section. 
Tests of the accuracy of our numerical procedures were extensively reported
in Ref.~\onlinecite{dv06} and need not be reported again here.
Different phases, corresponding to different vortex density 
structures, are 
obtained by starting the minimization 
process with different initial conditions.
For a new pin configuration, one can start either with uniform conditions
(all variables $\rho_{n,i}$ set to their average value) or with ``crystal
like'' initial conditions where the initial values of the  $\rho_{n,i}$
variables are set by minimizing\cite{prbv} the pinning energy 
of the equilibrium
crystal configuration in the absence of pinning. The first set of initial
conditions leads to a disordered configuration provided that such a 
configuration exists, at the values of $s$ and $T$ considered, as a local
free energy minimum, that is, either a stable or a metastable configuration.
Similarly, the second set of initial conditions leads to an ordered state
of the BrG type, if such a state is at least metastable. Once
a local minimum configuration has been obtained at a certain $s$ and $T$,
it can be used as the initial condition for a run at nearby values of these
parameters. This procedure is efficient since convergence is then fast
unless in changing the parameter values the boundary of stability
of the phase in question is crossed.
We have used this method to carry out 
various annealing and thermal cycling procedures~\cite{us2,prbd} 
that ensure that
the free-energy minima considered in our studies represent low-lying
local minima of the free energy.
Results for different random pin configurations
at the same values
of $c$ and $s$ are extremely consistent and 
we have found it sufficient to average
over three different configurations in order to obtain quantities
such as the location of phase boundaries.

\subsection{Structure of free-energy minima}

\label{struc} 

\begin{figure}
\includegraphics [scale=0.56] {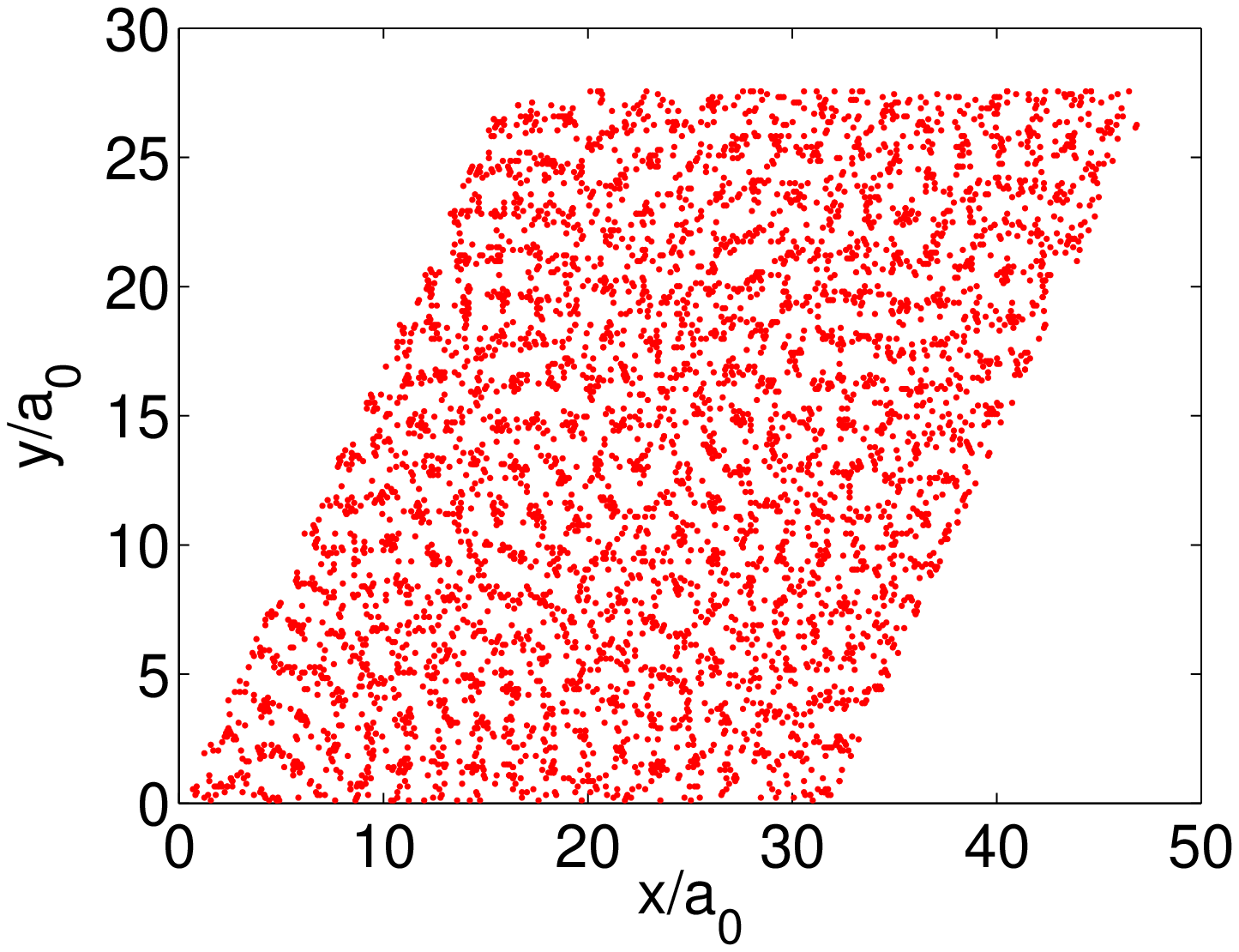}
\includegraphics [scale=0.56] {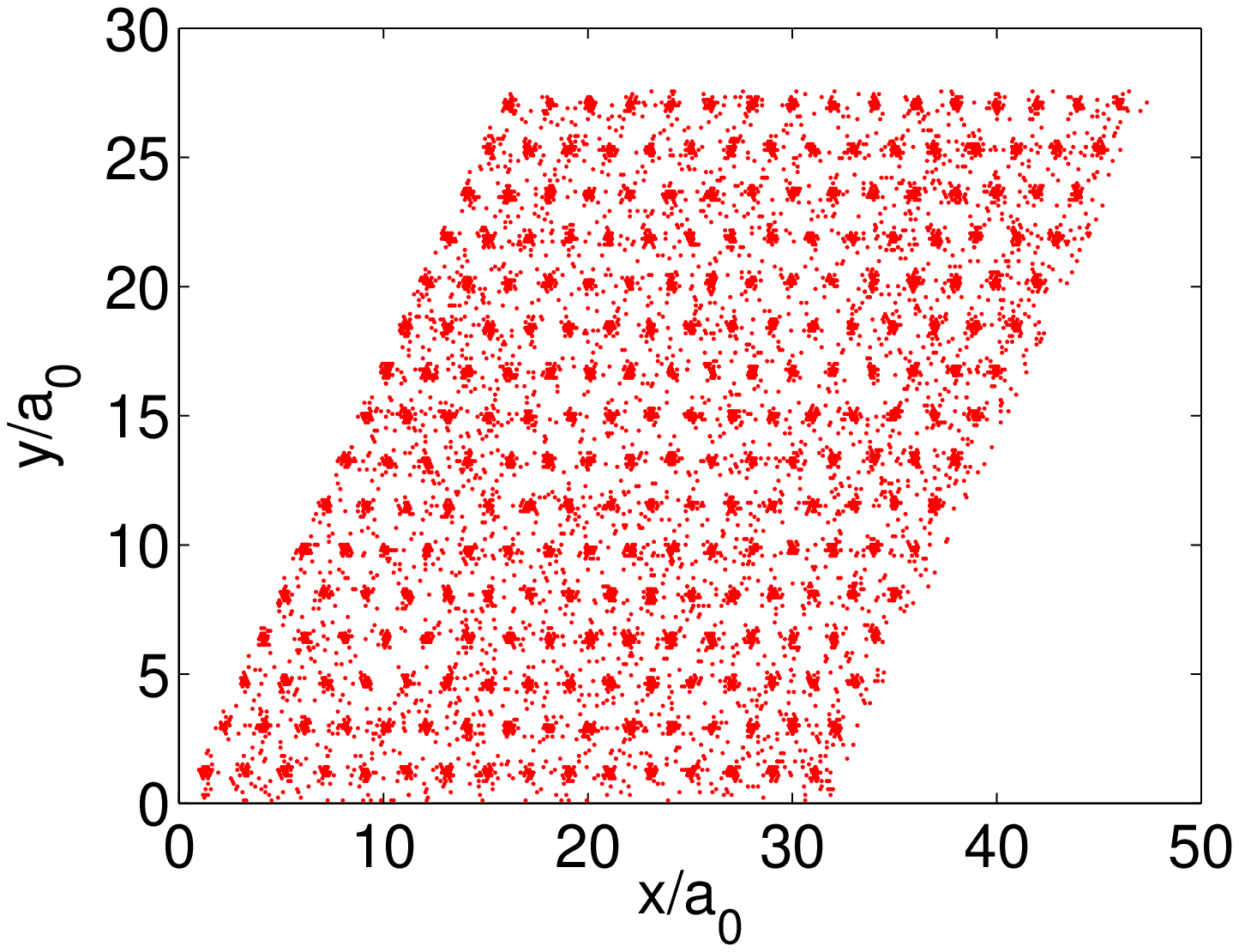}
\caption{(Color online) Peak positions at $s=120 K$,
$T=13.4 K$ for 20 consecutive layers in the disordered phase (top panel) 
and at
$s=120 K$, $T=15.2 K$ for all 128 layers in the ordered
phase (lower panel).
\label{fig1}}
\end{figure}

The structure of the free energy minima can be determined  by studying  
different correlation functions, that are readily obtained from
the  $\rho_{n,i}$ at that local minimum, and by direct visualization. 
In Fig~\ref{fig1} we consider the  option of directly visualizing
the {\it vortex lattice}. From the values of  $\rho_{n,i}$, we
extract the average vortex positions
by locating a vortex at site $i$ of layer $n$ if the value
of  $\rho_{n,i}$ at such a site exceeds all values of  $\rho_{n,j}$
for any site $j$ within radius $a/2$ of site $i$. In 
both panels of Fig~\ref{fig1}, a single
dot is plotted corresponding to the position of a vortex at any layer. In the
top panel which is for the disordered phase at 
$s=120 K$, $T=13.4 K$, this is done  for twenty consecutive layers,
chosen at random, while in the bottom panel, which is for the ordered phase at
$s=120 K$ and $T=15.2 K$, this is done for {\it all} 128 layers. The 
difference between the two phases can be seen clearly from these plots.
The ordered phase exhibits a nearly crystalline arrangement of the vortices,
while no such order is visible in the plot for the disordered phase.
If all 128 layers were included in the top panel, the plot
would be nearly filled by the dots. Thus, while there are, as we will see,
some interlayer correlations in the vortex positions in the disordered phase, 
their range is much shorter than that in
the ordered one.

\begin{figure}
\includegraphics [scale=0.8] {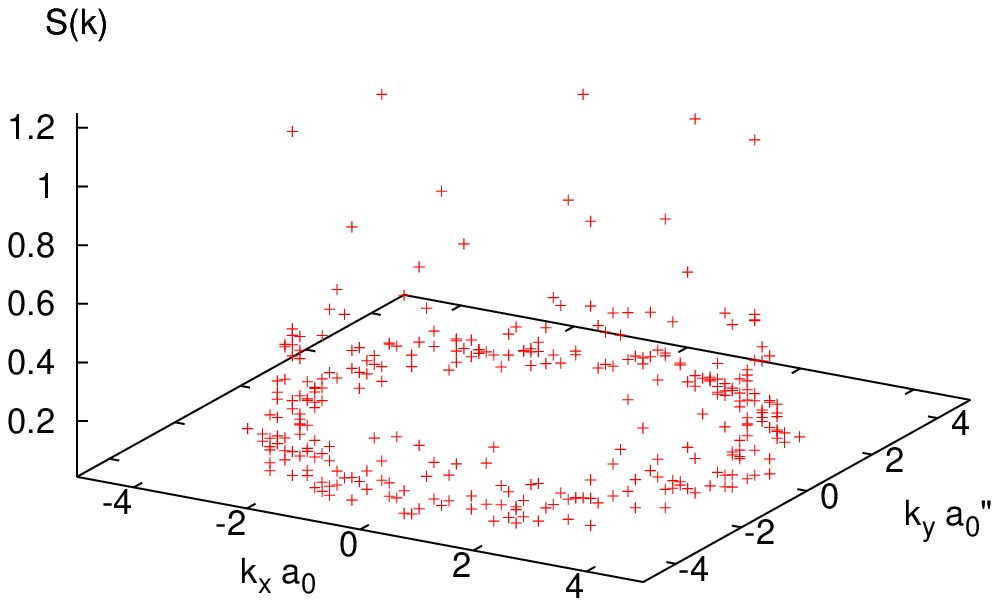}
\includegraphics [scale=0.8] {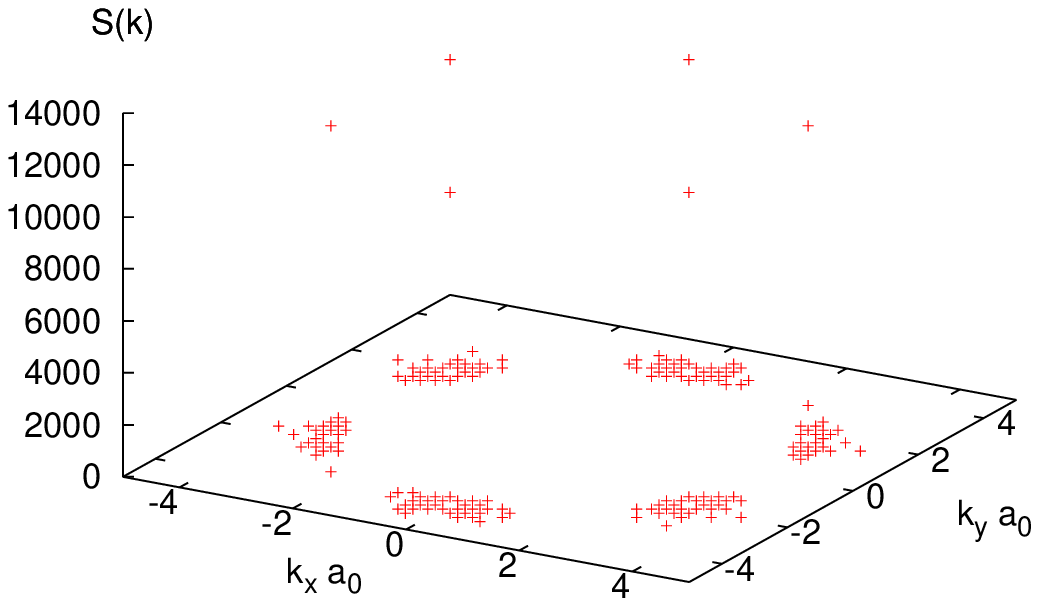}
\caption{(Color online) $S({\bf k})$ at $T=17.2 K$ and  $s=60 K$. Top
panel: disordered phase. Bottom panel: ordered phase.
\label{fig2}}
\end{figure}

Turning now to the correlation functions, we show in Fig.~\ref{fig2} the 
two-dimensional static structure factor $S({\bf k})$ defined as:
\begin{equation}
S({\bf k})=\frac{|\rho({\bf k},k_z=0)|^2}{N_v N_L},
\label{sk}
\end{equation}
where $\rho({\bf k},k_z)$ is the Fourier transform of $\rho_{n,i}$, with 
${\bf k}$ being the two-dimensional wave-vector in the layer plane and $k_z$
the wave-vector in the direction normal to the layers.
With the normalization
chosen (the total number of pancake vortices) $S$ should be of order unity
for a disordered state. Its maximum possible value is  $N_vN_L$ itself,
which equals 32768 in our samples.
Both panels of Fig~\ref{fig2} correspond to $s=60 K$, $T=17.2 K$ where both
phases are locally stable.
The top panel corresponds to the disordered phase and
the lower panel to the ordered phase which is the globally stable one, as we
will see below, at these values of $(T,s)$. Noting the very different vertical
scales and the well-defined hexagonal pattern in the lower panel, 
a clear difference between the two phases in the degree of order in
the transverse planes becomes quite obvious.
\begin{figure}
\includegraphics [scale=0.64] {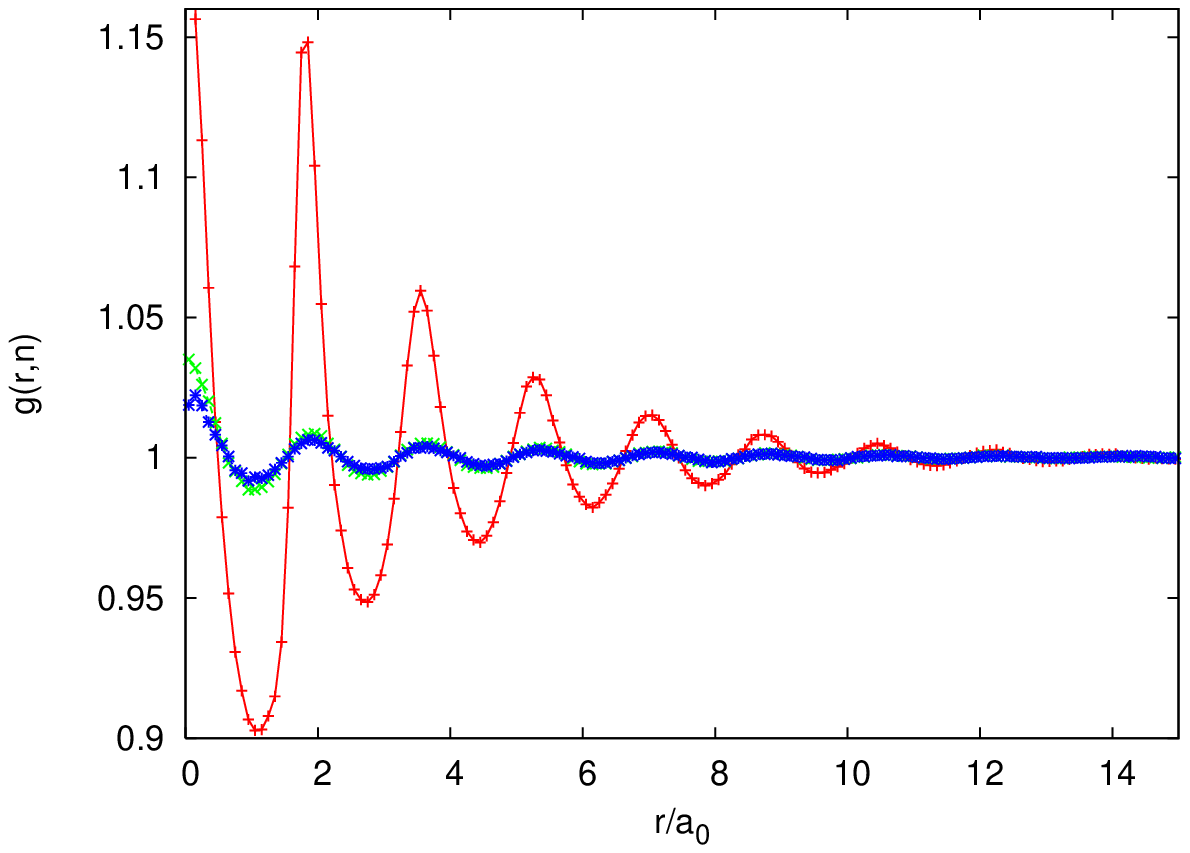}
\includegraphics [scale=0.64] {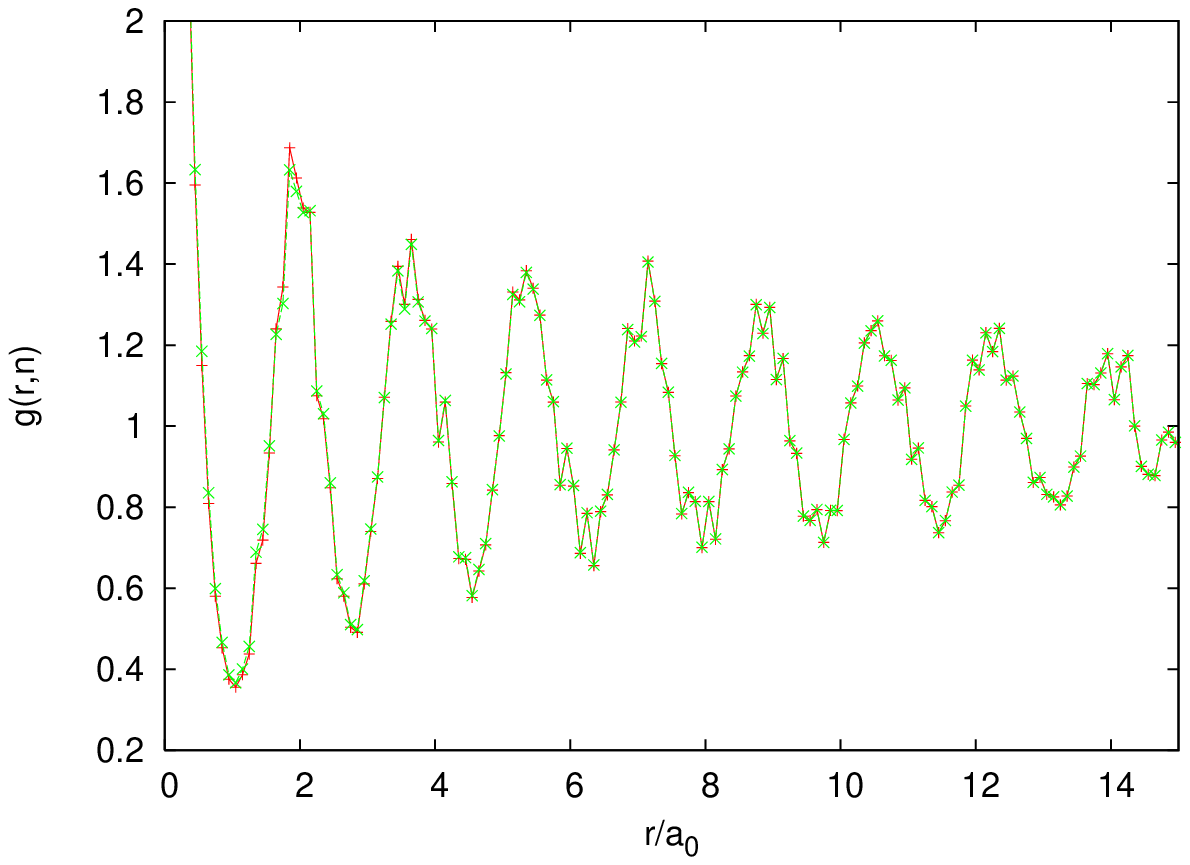}
\caption{(Color online) The 
real space correlations $g(r,n)$ as defined
in the text, 
at $s=100 K$,  $T=16.0 K$. The lines simply connect
the data points. Top
panel: results for the disordered phase for $n=0,1,9$ (from top to bottom
at $r=0$). 
Bottom panel: ordered phase, $n=0$ ((red) plus signs), $n=9$ ((green) crosses). 
\label{fig3}}
\end{figure}
This is confirmed by the results for the correlations
in real space, shown in Fig~\ref{fig3},  which pertain also to the
out-of-plane correlations. We define $g(r,n)$ as the 
angularly averaged Fourier transform
of $S({\bf k},k_z)$, normalized by $\rho_0^2$. 
Here $r$ is the in-plane distance and
$n$ indexes the distance between planes. 
Thus, $g(r,0)$ is simply the in-plane, angularly averaged correlation
function $g(r)$. We plot, in Fig~\ref{fig3}, $g(r,n)$ 
at $s=100 K$, $T=16.0 K$. At these values the ordered
phase is globally stable and the disordered one, metastable. In the top
panel (disordered phase) we plot 
the cases $n=0,1,9$. We see
that the decay of $g(r,n)$ as a function of either $r$ or 
$n$ is rather fast, particularly the latter. On the bottom panel, for
the ordered phase, where we plot only the two values $n=0,9$, the
situation is very different: the decay with $r$  is markedly slower but
the most remarkable feature is that the $n=0$ and $n=9$ results are nearly
identical, indicating long-range correlations in the direction normal
to the planes.

\begin{figure}
\includegraphics [scale=0.56] {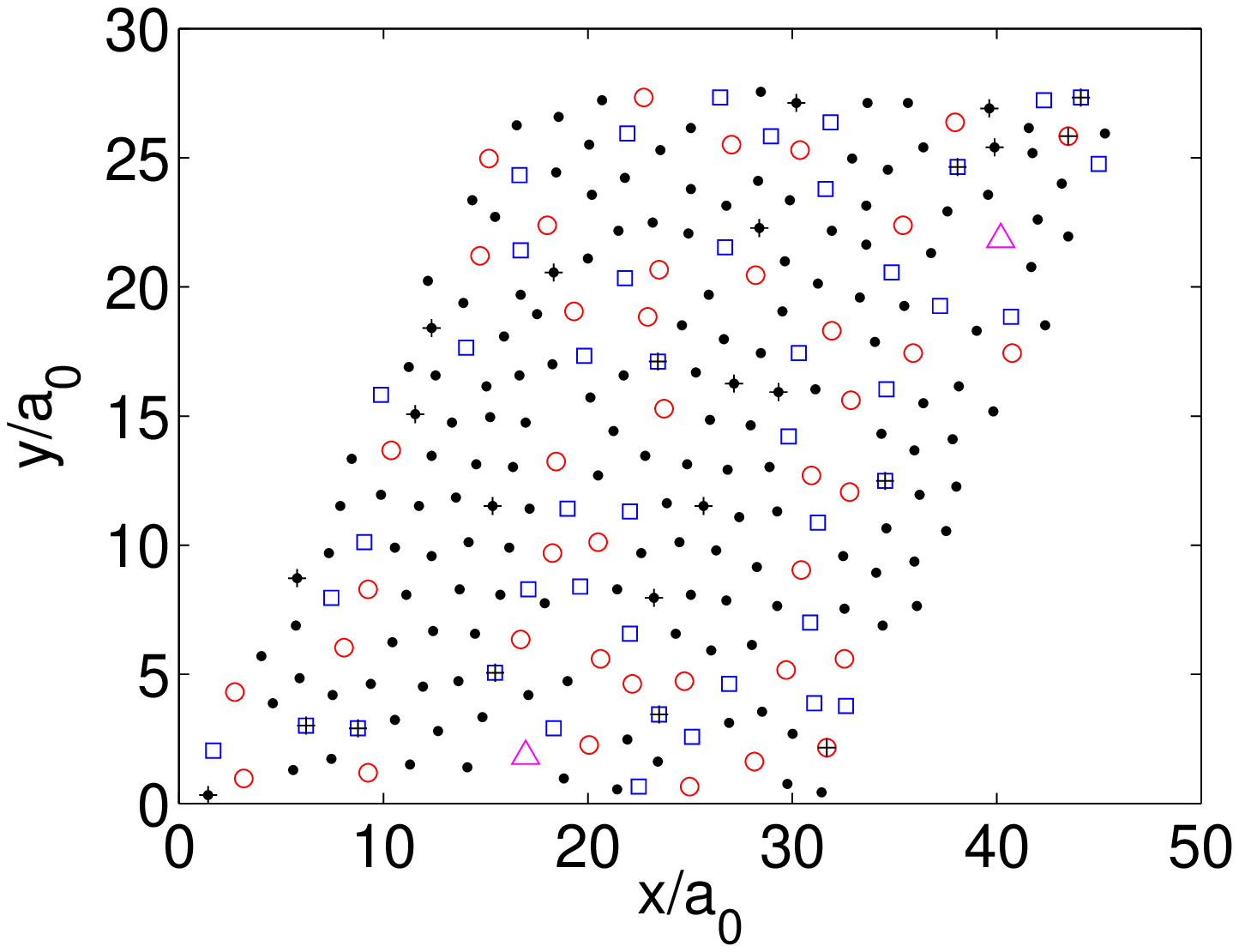}
\includegraphics [scale=0.56] {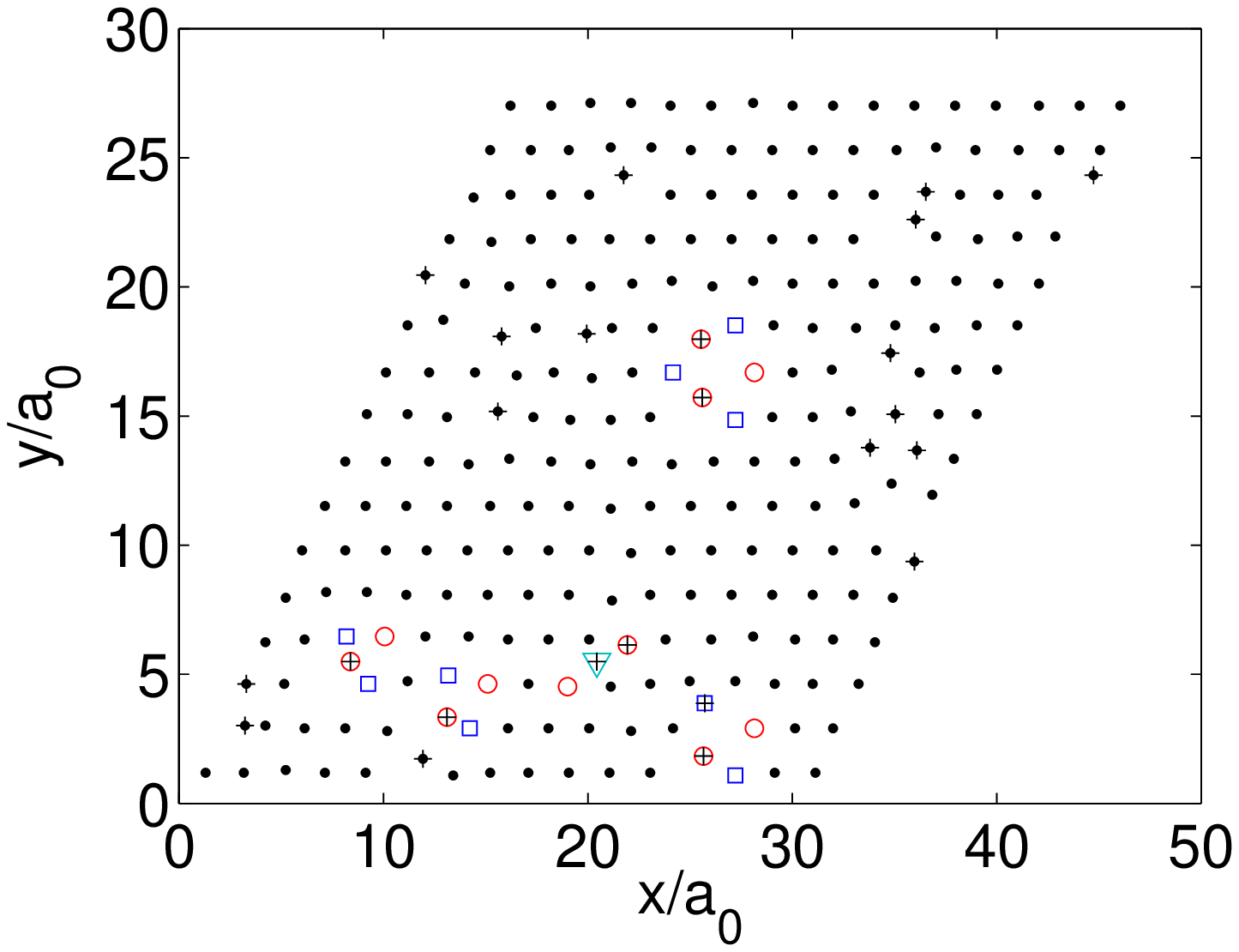}
\caption{(Color online) Voronoi plots.  The (black) dots
denote six-neighbor sites, the (red) circles denote sites with seven
neighbors, (blue) squares sites with five neighbors, 
and (magenta) triangles sites with eight neighbors. 
The (green) inverted triangle in the bottom panel represents a site with 
four neighbors. The (black) plus signs 
locate pinning sites. Top and bottom panels are for the same configurations
as in Fig.~\ref{fig1}.
\label{fig4}}
\end{figure}

We now return to the vortex lattice, as in Fig.~\ref{fig1}, and re-examine
it via a Voronoi construction. We recall that the Voronoi construction is
nothing but the usual Wigner-Seitz procedure 
on a perfect lattice, but performed in an arbitrary network.  
The number of sides of a Voronoi cell associated with a vertex of
the network (location of a vortex in our case) represents the number
of neighbors of that vertex. These numbers of neighbors are represent
by a symbol (and color) code in Figure~\ref{fig4}.  There we display
Voronoi plots for one layer, chosen at random, of the systems whose
vortex configurations are shown in Fig.~\ref{fig1}. One can see that in
the ordered phase (bottom panel) nearly all the vortices have six
neighbors and that the most common deviation is a pair of
adjacent sites with five and seven neighbors, which corresponds to a 
dislocation. These dislocations invariably occur near pinning sites (in fact,
all the visible deviations from a perfect crystalline arrangement of the 
vortices are found near pinning sites which are almost always occupied by
vortices, except in very rare occurrences of two pinning sites very close
to each other), and they form tightly bound
clusters with net Burgers vector equal to zero.
The top plot is very different: there we can see nearly ordered
domains in which the vortices have the regular number of six neighbors. These
domains are small and separated by regions 
that are  much more disordered, consisting mainly of sites with five and
seven neighbors. These regions might be called grain boundaries if they
were uniformly thinner.

>From these plots and similar data obtained in the ranges $60K \leq s 
\leq 170 K$, 
$13.4K <T <19.0 K$, we conclude that we have here 
two different phases, separated
by a first-order transition. The evidence given up to this point suffices
to show that the ordered phase is a Bragg glass. The
nature of the disordered phase requires more detailed characterization,
which is addressed in the following subsection.

\subsection{Vortex slush and pinned vortex liquid}
\label{slush}

In this section, we present the results of a detailed characterization of the
structure of the disordered phase. We compare and contrast 
these characterization
results with those obtained for the disordered phase 
in our earlier study~\cite{dv06} of the
structure of the disordered phase of the same vortex system in the
presence of a {\it much larger
} ($1\%$ of the underlying crystal lattice)
concentration $c$ of randomly placed {\it weak} (small $s$)
pinning centers. To make this comparison
meaningful, we particularly
consider cases where despite the differences in the values
of $s$ and $c$, the values of the
parameter $\beta \sqrt{s^2c}$ in 
the two cases are comparable. As explained 
above, if the
effects of random pinning were determined entirely by the value of $s^2c$,
then the structures of the disordered phases in the two systems would have been
very similar. As we will seen in detail, we find very significant 
differences between the structures of the two disordered phases: indeed,
we show that while the disordered phase in the system with a large 
concentration of weak pinning centers has very little short-range order
(this led us to conclude~\cite{dv06} 
that this phase should be identified as a pinned
vortex liquid), the disordered phase in the present system with a small
concentration of strong pinning centers exhibits substantial short-range
order, both in the layer plane and in the direction normal to the
layers, leading us to identify this phase as VS or MG.

\begin{figure}
\includegraphics [scale=0.56] {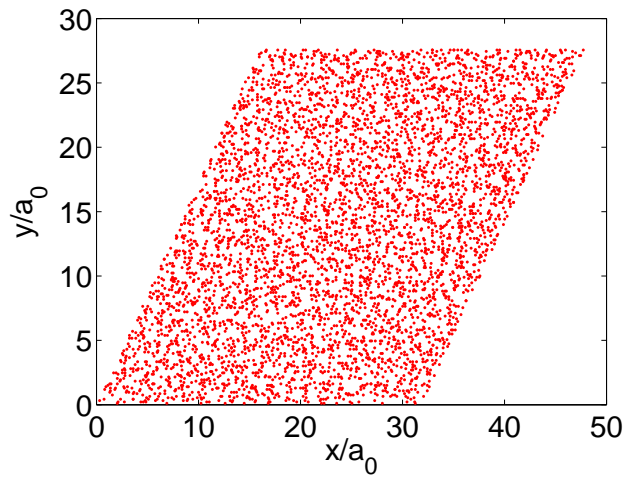}
\caption{(Color online) Peak positions  of twenty randomly
chosen consecutive layers at $T=18.6 K$
and a large concentration ($1\%$ of the underlying crystal lattice)
of weak ($s=4K$) point defects, as in Ref.~\onlinecite{dv06}. Compare
with Fig.~\ref{fig1}, top panel.
\label{fig5}}
\end{figure}

\begin{figure}
\includegraphics [scale=0.56] {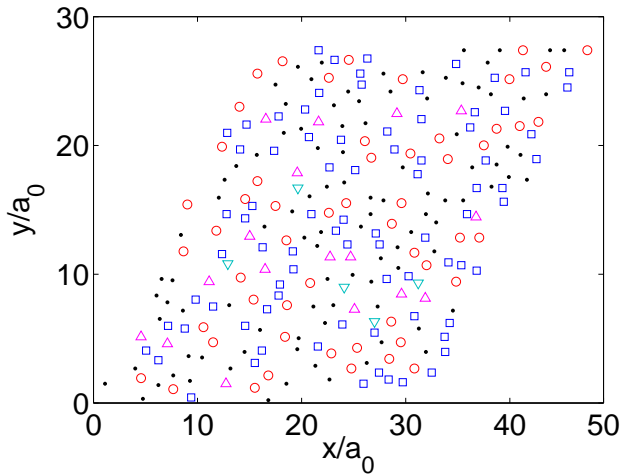}
\caption{(Color online) Voronoi plot at $T=18.6 K$
and a large concentration ($1\%$ of the underlying crystal lattice)
of weak ($s=4K$) point defects, as in Ref.~\onlinecite{dv06}.  The
meaning of the symbols is the same as in 
the top panel in Fig~\ref{fig4}, with which the current plot should
be compared.  
\label{fig6}}
\end{figure}

We begin with a qualitative comparison  of the peak-position plots  
(see Fig.~\ref{fig1},
top panel), and of the Voronoi plots (see Fig~\ref{fig4}, top panel). In 
Fig.~\ref{fig5}, we have shown the positions of the local density peaks
on randomly selected 20 consecutive layers of a disordered free-energy
minimum obtained at $T=18.6$K for a sample\cite{dv06} with the larger
value of $c$ and 
weak ($s=4$K) pinning centers. The dots that represent the
peak positions fill the sample area quite randomly, in contrast to the
plot shown in the top panel of Fig.~\ref{fig1}, which shows a definite
short-range structure,
indicating the presence of correlations among the positions of the local
density peaks. Similarly, a typical Voronoi plot 
for a randomly chosen layer of the
minimum for which 
results are given in Fig.~\ref{fig5} is shown
in Fig.~\ref{fig6}. Again, in comparison with the analogous
plot shown in the top
panel of Fig.~\ref{fig4}, this Voronoi plot shows a much higher degree of
disorder. In particular, the total number of ``vortex lattice'' sites with 
six nearest neighbors is substantially smaller in Fig.~\ref{fig5}, and 
there is no evidence for the presence of any crystalline domain (for this
reason, we classified~\cite{dv06} the structure of these minima 
as amorphous, not
polycrystalline, in our earlier study). These two plots, therefore, suggest 
that the disordered minima obtained in the present study exhibit a much higher
degree of short-range order than those found earlier for a system with a large 
concentration of weak pinning centers.

To make these observations
more quantitative, we have calculated two correlation
functions, $C(r)$ and $C_z(n)$, that represent, respectively, the degree
of correlation among the {\it vortex positions} (i.e. the positions of the
local peaks of the density field) in the layer plane and across layers. The
function $C(r)$ is the usual angularly averaged 
pair-distribution function of vortex positions
on the same layer, averaged over all the layers. It should not
be confused with the
correlation function $g(r) \equiv g(r,n=0)$ considered in section~\ref{struc}
which represents the two-point correlation of the  time-averaged {\it local
density}: therefore, $g(r)$ is sensitive to both the positions  
and the heights of the
local density peaks, whereas $C(r)$ provides more accurate information about
the correlation in the peak positions because it is insensitive to the heights
of the local density peaks. The function $C_z(n)$ is defined as follows:
For each peak position on a layer, we calculate the 
number of peaks on another layer, separated by $n$ layer spacings from
the original one, that lie within a small in-plane
distance $r_0$ from it. This number is then averaged over all the 
peak positions on all the layers and divided by $\pi r_0^2 \rho_0$, 
the average number of vortices in a circular area of radius $r_0$, to
obtain $C_z(n)$. The values of $C_z(n)$ are found to be insensitive to the
choice of $r_0$ as long as $r_0$ is small compared to the average distance
between nearest-neighbor vortices ($\simeq 1.9a_0$, see Fig.~\ref{fig3},
top panel) on the same layer. The results reported here were obtained using
$r_0 = 0.22 a_0$. It is clear from the definition of $C_z(n)$ that this
function measures the degree of alignment between vortices on two layers
separated by $n$ layer spacings. 
If the vortex positions on different layers are 
completely uncorrelated, then $C_z(n)$ should be equal to unity for all
$n$. Values of $C_z(n) > 1$ indicate some degree of alignment between the
vortex positions on different layers.

\begin{figure}
\includegraphics [scale=0.56] {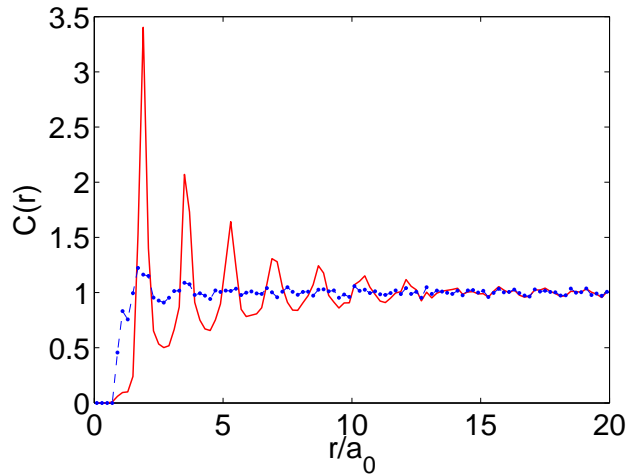}
\caption{(Color online) The in-plane vortex correlation function
$C(r)$ (see text) as a function of $r$, for the disordered minimum
obtained in the present calculation at $s=170 K$, $T=14.6 K$ 
((red) solid line),
compared with the result for weak point defects shown in Figure~\ref{fig6}
((blue) dotted line).
\label{fig7}}
\end{figure}

\begin{figure}
\includegraphics [scale=0.56] {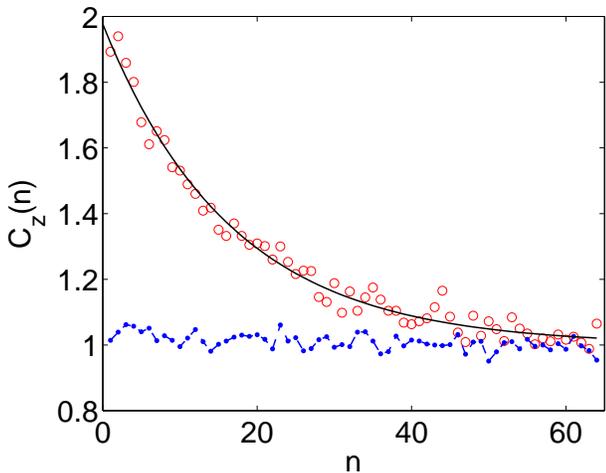}
\caption{(Color online) The out-of-plane vortex correlation function
$C_z(n)$ (see text) shown for the same two cases as in the preceding figure.
The (red) circles are the high $s$ case of the present work. The solid
line is an exponential fit. 
\label{fig8}}
\end{figure}

Typical results for these two correlation functions are shown in 
Figs.~\ref{fig7} and \ref{fig8}. The results for a system with a small 
concentration of strong pinning centers (present study) are shown for
$s=170$K and $T=14.6$K, whereas the results for a system with a large
concentration of weak pinning centers (Ref.~\onlinecite{dv06}) are for the
same parameters as in Figs.~\ref{fig5} and \ref{fig6}. These 
were chosen so as to make the values of $\beta \sqrt{s^2c}$ for the two
systems very similar. Also, the functions $C(r)$
and $C_z(n)$ that describe correlations among average 
vortex positions are meaningful
only when the local density peaks are sufficiently high (the vortices
are strongly localized), so that the average 
vortex positions can be defined without
ambiguity. This criterion is satisfied for the parameter values for which 
results are shown in Figs.~\ref{fig7} and \ref{fig8}.

The plots in Fig.~\ref{fig7} clearly show that the disordered phase in the
present (low $c$) system exhibits a substantial degree of in-plane 
(short-range) positional order,
while the disordered phase in the system with a large concentration of
weak pinning centers has almost no in-plane correlation in the vortex
positions. The function $C(r)$ for this low $c$ and high $s$ system 
exhibits six clear peaks of decreasing height, indicating the
existence of in-plane positional correlations up to a distance of about
$10a_0$. This length scale is consistent with the typical size
(20--30 vortices, see Fig.~\ref{fig4}, top panel) of in-plane crystalline
domains at the disordered free-energy minima found in the present work.  
In contrast, the $C(r)$ for the system with a large concentration of weak
pinning centers exhibits no prominent peaks, indicating that the 
disordered phase of this system has very little in-plane positional
order and crystalline domains can not be meaningfully defined in this case.

A similar difference in the degree of out-of-plane correlations in the vortex
positions is also found in Fig.~\ref{fig8} where the function $C_z(n)$ is
plotted for the two free-energy minima of Fig.~\ref{fig7}. The data for the
system with a large concentration of weak pinning centers are all close to
1.0, even for values of $n$ near 1. This implies that the vortex
positions on different layers are essentially uncorrelated in this case. On
the other hand, $C_z(n)$ for the present system starts at a value close to
2.0 for $n$ near 1, and approaches unity
only  as $n$ approaches about $N_L/3$,
(the largest value of $n$ for which $C_z(n)$ can be defined in our system
of $N_L$ layers with periodic boundary conditions is $N_L/2$). 
These results indicate that
the vortices on different layers are aligned to some extent in the 
disordered phase of the present
system. We find that the dependence of $C_z(n)$ on the layer separation $n$
can be represented quite well by the functional form $C_z(n) = 1.0 +
C_{z0} \exp(-n/l_z)$ where the ``correlation length'' $l_z$ provides a
measure of the degree of out-of-plane alignment of the vortices. A fit of
the data to this functional form (with $C_{z0}=1.0$, $l_z=17.2$) is shown
by the solid line in Fig.~\ref{fig8}. Similar values of $l_z$ are found for
other values of the parameters in the disordered phase 
of the present system, indicating that the vortex positions on different
layers are substantially correlated in the system with a small concentration
of strong pinning centers.

It is clear from all the results described above that the disordered 
free-energy minima found in the present study exhibit a substantial degree
of short-range positional order. The in-plane structure can be described as
polycrystalline, with 20-30 vortices in each crystalline domain. The 
vortex positions on different layers are also correlated, with a correlation
length of 15-20 layer spacings. These features are similar to those
expected for the VS and MG phases that have been
predicted to exist in vortex systems with random point pinning. We, therefore,
identify the disordered phase in the present system as VS. This
is qualitatively different from the disordered phase found in our earlier
study~\cite{dv06} of a vortex system with a large concentration of weak
pinning centers. In that case, the disordered phase, which does not exhibit 
any appreciable in-plane or out-of-plane correlation in the 
vortex positions, was identified as pinned vortex liquid which should be
distinguished from the VS found in the present work.

\subsection{Phase Diagram}
\label{phdiag} 

\begin{figure}
\includegraphics [scale=1.0] {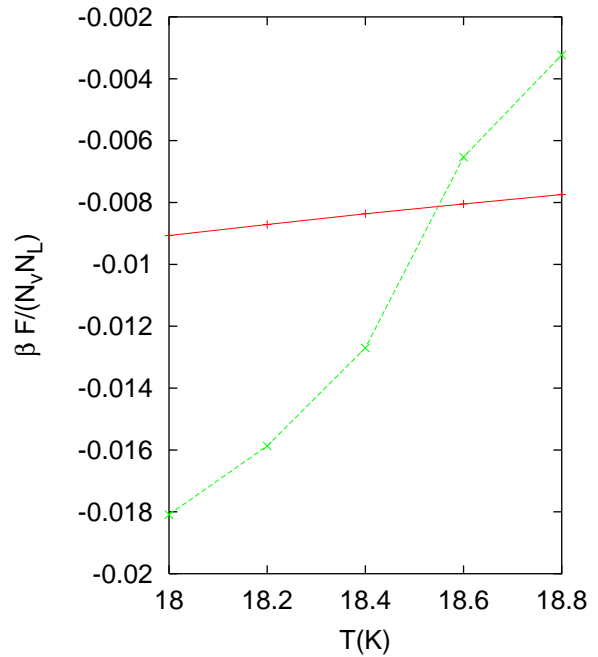}
\caption{(Color online) Free energy crossings
at $s=60 K$ as a function of $T$. The quantity plotted is the free energy
per vortex in units of $k_BT$. The (red) plus signs are results  for the 
VS phase and the (green) $\times$ signs for the BrG phase. Straight
lines
connect the data points. 
\label{fig9}}
\end{figure}

To determine the phase diagram one simply looks for free energy crossings
as a function of $s$ and $T$. This is quite straightforward. 
An example is shown
in Figure~\ref{fig9}, where we consider the free energy per vortex, in units
of $k_BT$, for the BrG and VS phases 
at constant $s$ as a function of temperature. 
In the range plotted both phases
are locally stable. The BrG minimum becomes unstable 
just beyond the right edge of
the plot. The symbols correspond
to the results for the free energy and they are connected
by straight lines.
Error bars from statistical averaging over different pin configurations
are smaller than the symbols. The first order transition between the two phases
is easily located from such data. A similar first order 
transition from the BrG to the VS
phase is  found as the pinning strength $s$  
is increased while keeping $T$ 
constant at a low value. The BrG minimum becomes unstable 
as $s$ or $T$ is increased
slightly above the value at which the transition to the VS 
phase occurs. The VS minimum,
on the other hand, remains locally stable for all the values 
of $s$ and $T$ considered here.

For relatively small values of $s$ and 
large $T$, the values of the density at the local peaks 
of the density field in the VS phase do not
greatly exceed (see below) the average liquid density value $\rho_0$
except for
peaks at the pinning centers. In that sense, the VS phase might
be said to be liquid-like.
On the other hand, for large $s$ and 
low $T$, these local peak density values 
are substantially larger than $\rho_0$, 
indicating strongly localized
vortices. A VS minimum with these properties should be 
viewed as solid-like (glass).
However, for the parameter range explored 
here, we do not find
any evidence for a transition within the VS state between the high $T$-low $s$
and the low $T$-high $s$ structures. A liquid-like minimum 
obtained by starting the free-energy
minimization from disordered initial conditions 
at a high $T$ evolves continuously
to a glassy VS minimum as it is ``followed'' to a 
low temperature by reducing $T$ in small
steps and performing the free-energy minimization at 
each new temperature with the minimum
obtained at the previous temperature as the initial state. 
Similarly, a glassy VS minimum obtained
at low $T$ and large $s$ by starting the minimization 
from disordered initial conditions (or by 
starting from a BrG minimum and increasing $s$ to a value 
at which the BrG minimum becomes
unstable and a
disordered minimum is found) evolves continuously to the 
high-temperature liquid-like structure as $T$ 
is increased in small steps. 

Thus,
our results do not show the first-order VL to VS transition 
found in some
experiments~\cite{slush1,slush2,slush3} and in a simulation~\cite{nh}. 
However, these experiments and simulation
show that the line of first-order 
VL to VS transitions ends at a critical point as the
pinning strength (in the simulation) or the magnetic field 
(in the experiments) is increased. 
It is likely that the parameters
used in our study correspond to values beyond this 
critical point, so that no VL to VS
transition is found. In a study~\cite{us1,us2,prbd} 
of the same vortex system in the
presence of a small concentration of strong columnar pins 
(with values of $s$ comparable to
those used here) perpendicular to the layers, we found a 
first-order transition from the
high-temperature VL to a polycrystalline Bose glass phase 
as $T$ was decreased, but only 
when the ratio of the number of pinning centers to the 
number of vortices per layer was less than 
about 1/32. The value of this ratio  
in the present study (24/256 = 3/32) is larger than 1/32. 
This suggests that a first-order VL to VS
transition may occur in the present system for a smaller 
concentration of pinning centers.
It is, however, difficult to test this possibility 
numerically. The typical
size of crystalline domains in a VS state will 
increase as the concentration of pinning centers is
reduced. A multi-domain structure can be distinguished from a crystalline one 
in a numerical study only if the
sample size is substantially larger than the 
typical domain size. So, to study polycrystalline free-energy
minima in the present system for a much smaller 
concentration of pinning centers, one 
would require samples containing a few thousand
vortices per layer. Samples of such size could be studied 
for columnar pins 
because the problem is then effectively two-dimensional, but
a similar study for the
present case (where the problem is
three-dimensional) would be computationally intractable. 

\begin{figure}
\includegraphics [scale=0.56] {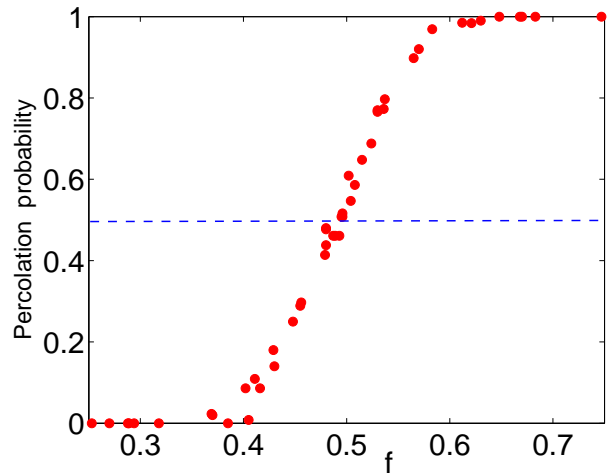}
\caption{ (Color online) Percolation probability plotted 
as a function of the fraction
$f$ of liquid like sites.  The (red) dots are the data, as explained
in the text. The (blue) dashed line is along the value of 1/2.
\label{fig10}}
\end{figure}

A crossover line between liquid-like and glassy 
behaviors of the VS phase can  be drawn 
in the $(T-s)$ plane  by using a criterion
based on percolation. 
In Ref.~\onlinecite{prbd}, we found that a local peak of the density field may
be classified as ``liquid-like'' (the corresponding 
vortex is weakly localized) 
if $\rho_{peak}$, the value of the density variable at the peak, does 
not exceed $3\rho_0$. Otherwise it is characterized as ``solid-like'', 
representing a strongly localized vortex. One can then
readily examine whether or not, at a certain $s$ and $T$, the 
regions with $\rho_{peak} < 3 \rho_0$ percolate across any 
of the layers in a sample. If such regions percolate in a 
majority of the layers, then the corresponding VS minimum may be 
classified as ``liquid-like''; otherwise, the minimum should 
be called ``glassy''.
Similar percolation criteria
have been used in other studies~\cite{perc1,perc2} to
differentiate between liquid and glassy phases. 
We have done this for all values of $s$
and $T$ studied. The probability
that liquid-like sites percolate across a layer depends on the fraction
$f$ of such sites. This dependence 
is shown in Fig.~\ref{fig10},
where we see that the percolation probability is 1/2 very close
to $f=1/2$. We have therefore adopted the criterion that percolation
of liquid-like regions occurs at $f=1/2$, which is also  
the percolation threshold
for random site percolation on a triangular lattice.

\begin{figure}
\includegraphics [scale=1.0] {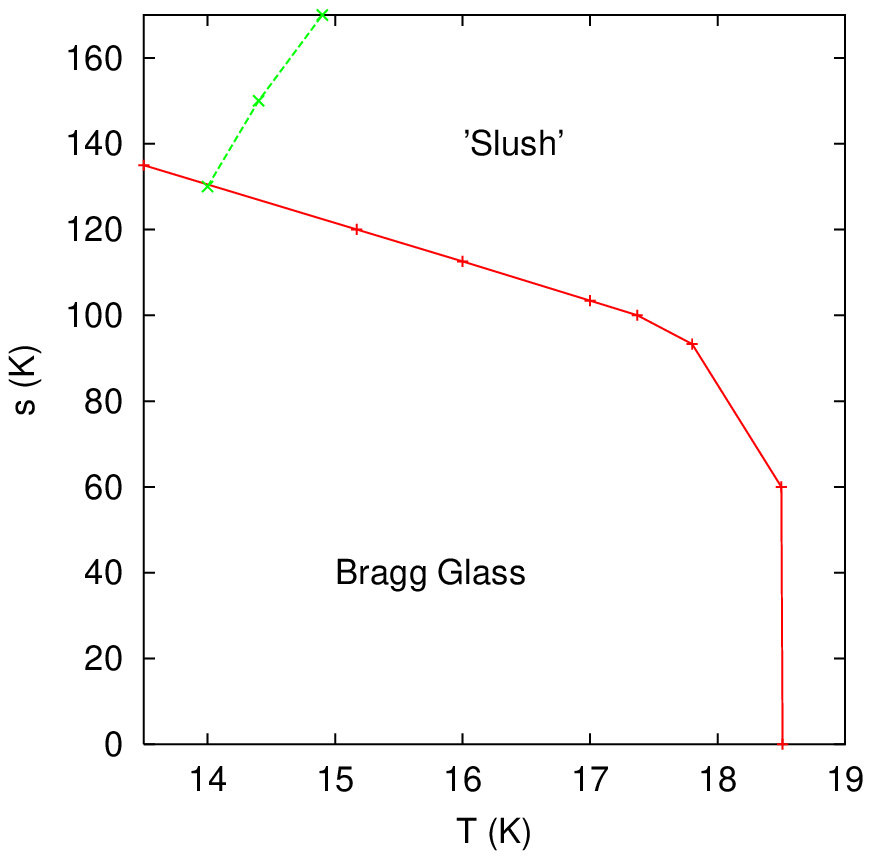}
\includegraphics [scale=1.0] {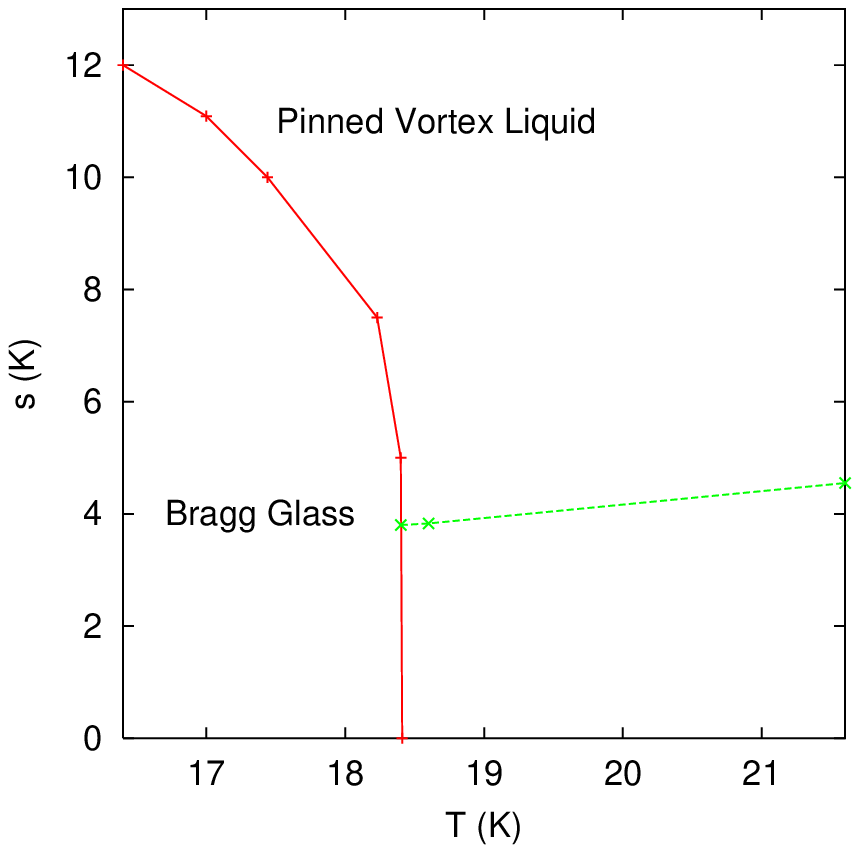}
\caption{\label{phdgm} (Color online) Phase diagram (top panel)
for the present case. The (red) plus signs are the results
for the first order transition between the BrG and 
VS phases. The points are connected by straight
line segments. The (green) $\times$ signs connected
by a dotted line denote the points at which percolation of liquid-like
region occurs in the VS phase. 
For comparison, the bottom panel shows the phase diagram
at a high concentration of weak pinning centers (Ref.~\onlinecite{dv06})
including the percolation line.
\label{fig11}}
\end{figure}

The phase diagram of the system in the $(T-s)$ plane is shown 
in Fig.~\ref{fig11}, top panel. The plus signs 
connected by solid red segments locate the  line
of first order transitions separating, as explained above, 
the BrG and the 
VS phases. We have also indicated in the panel the percolation
line for high $s$ and low $T$ -- percolation of liquid-like regions
in the VS phase occurs 
as this line is crossed from the left or from above. A few features of 
the phase diagram are worth pointing out. The temperature at which the 
melting transition of the BrG phase occurs is quite insensitive to the
value of $s$ for relatively small values of $s$. As $s$ is increased further, 
the transition temperature decreases and the phase boundary 
tends to become parallel to the 
temperature axis for large $s$. The decrease in the transition temperature 
with increasing $s$ is in qualitative agreement with the prediction of analytic
calculations.\cite{gautamprl,goldsc,gld2} 
The BrG phase exists only if the pinning strength is sufficiently small.
The shape of the BrG--VS phase boundary in our phase diagram is
similar to that 
found in the simulation of Ref.~\onlinecite{nh} in which the ratio 
of the number
of pinning centers to the number of vortex lines is similar to the value
used here. 
The location of the percolation line relative 
to the BrG--VS phase boundary is also similar to that of the VS--VG transition
line found in Ref.~\onlinecite{nh}. This, 
however, may be a coincidence, since the percolation line in our phase
diagram does not correspond to a true phase transition, whereas the VS--VG
transition line in Ref.~\onlinecite{nh} represents a phase transition 
signaled by the appearance of superconducting phase coherence which can not 
be studied in our calculation.

For comparison we have included in
Fig.~\ref{fig11} (lower panel) the phase diagram obtained\cite{dv06} 
for a high concentration of weak pinning centers, 
including also the percolation line, which was not
reported earlier. The general shapes of the phase boundaries obtained in the
two cases are similar. However, a quantitative comparison shows that 
a low concentration of strong pinning centers considered in the present
study is much more effective in destroying the nearly crystalline
order of the BrG phase than a large concentration of weak pinning centers
considered previously. For example, in the phase diagram 
shown in the top panel of Fig.~\ref{fig11}, the transition of the BrG phase
to the VS phase at $T=17 K$ occurs at $s \simeq 100 K$, for which the rms
pinning potential is $\sqrt{268.3 s^2 c} = 1.9 K$ (see Section~\ref{methods}).
In contrast, the transition at $T=17K$ in the phase diagram in the bottom
panel of Fig.~\ref{fig11} occurs near $s=11 K$, for which the rms value of the
pinning potential is $18.0 K$. Thus, the rms value of the random
pinning potential at which the BrG phase undergoes a transition to a 
disordered phase at a fixed temperature in a system
with a large concentration of weak pinning centers is about 
{\it 10 times larger}
than that in a system with a small concentration of strong pinning centers.
This implies that the kind of pinning considered in the present study is much
``stronger'' (i.e. more effective in destroying BrG order) 
than that considered in Ref.~\onlinecite{dv06}.

The comparison between the two phase diagrams 
also points out a serious flaw in analytic calculations in
which the random pinning potential is assumed to be Gaussian and its effects
are assumed to be determined completely by its rms value (second moment).
Our results show that the details of the pinning potential, not just its second
moment, are very important in determining the phase behavior: the value
of the second moment of the pinning potential at which the BrG phase undergoes
a transition to a disordered phase at a fixed temperature can vary by 
orders of magnitude, depending on the details of the pinning potential.
Thus, the predictions of analytic 
calculations~\cite{gautamprl,goldsc,gld2,nelson,larkin} 
of the phase diagram of superconductors with random pinning, in which
the effects of the pinning potential are assumed to be determined by its
second moment only, can not be quantitatively accurate -- the details of
the pinning potential have to be taken into account for an accurate 
theoretical determination of the locations of the phase boundaries. 

Fig.~\ref{fig11} also shows that the location of the percolation line 
relative to that of the transition line of the BrG phase is very different
in the two cases. In the top panel, the percolation line intersects the
BrG--VS transition line at a relatively low temperature 
where the latter is almost parallel to the $T$-axis,
whereas in the bottom panel, the intersection is near the 
high-temperature, ``vertical'' part
of the boundary of the BrG phase. The value of the dimensionless quantity
$\beta \sqrt{s^2 c}$, which measures the ratio of the rms value of the pinning
potential to the thermal energy is, along the percolation line, 
nearly the same
in the two cases, indicating that the rms value of the pinning potential
provides a fairly accurate account of the effectiveness of random pinning in
localizing the vortices. The difference in the location of the percolation
line relative to the phase boundary of the BrG phase  again arises 
because the  rms value of the pinning potential is {\it not},
by itself,  a good indicator
of the effectiveness of the pinning potential in destroying BrG order.

\section{Summary and discussion}
\label{summary}

We have considered
here a layered superconductor with a small concentration $c$ of
strong random  point pinning centers and
a fixed magnetic induction, normal to the layers. We have described the
phase diagram in the $(T-s)$ plane, where $s$ is 
the pinning strength. We find that the stable phase
at sufficiently low  values of $T$ and $s$ is clearly a Bragg glass (BrG),
as in the high $c$ case.\cite{dv06} 
The melting line in the $(T-s)$ plane at low $T$ is nearly horizontal.
Upon increasing $s$ or $T$ the BrG melts via a first order transition
into a disordered phase.
We find that this phase is clearly polycrystalline
(see top panels in Figs~\ref{fig1} and \ref{fig4}) 
with well defined crystalline domains separated by domain walls.
It has characteristics of the VS phase discussed in the  
literature,\cite{nh,slush_th1,slush_th2,slush1,slush2,slush3} and
of the also proposed\cite{banerjee,gautam} MG phase.
We therefore characterize this phase (see Fig.~\ref{fig11} and
the text) as a VS.
There is a very clear difference  between this and the high $c$, low $s$ 
case as to the
nature of this disordered phase.
For large $c$ and low $s$, this phase 
was found\cite{dv06} to be an amorphous, pinned VL state, 
with almost no correlations in the layer plane
or between different layers
(see Figs.~\ref{fig6} and \ref{fig7}). 
A key result of this paper is this contrast  between the behavior of the
system studied here  and that of 
the corresponding system\cite{dv06} with a  
much larger concentration of weak pinning centers.

The phase diagrams of these two systems in the $(T-s)$ plane are qualitatively
similar, but very different quantitatively. The value of the 
second moment of the random pinning potential at which the transition of the
BrG to the VS occurs in the present system is much smaller than that at the
BrG to pinned VL transition in Ref.~\onlinecite{dv06} at the same temperature.
Also, the location of the percolation line relative to the phase boundary
of the BrG phase is quite different in the two systems.
These results imply that the notion that the parameter $\beta \sqrt{s^2c}$
suffices  to characterize the effects of the random point 
pinning potential (in
other words, that it is enough to specify its second moment)
is too
simplistic and may be quantitatively inaccurate.  
The assumption of Gaussian randomness,
which implies that only the value of the second moment of the distribution
of the random pinning potential is relevant, 
is valid only in a rather limited sense, namely 
only when the pinning potential seen by a vortex is the sum of contributions
from a large number of weak pinning centers. The 
validity of the Gaussian assumption does
not  extend further than that, as we have proved here. 
Although the limited validity
 of this assumption
was mentioned in a few 
existing studies,\cite{gautamprl,goldsc}
 which did not however quantify
its degree of inadequacy, 
it has been generally
ignored.


Our
results  suggest that many of the 
experimental and theoretical controversies found in the literature
(which we have reviewed in our Introduction)  about
the behavior of this system originate in differences in 
the insufficiently characterized  nature of the
point pinning. This conclusion, besides its obvious theoretical
import, has also strong practical implications since a good understanding
of the effects of pinning is the key 
to obtaining the larger values of the critical
current that are needed for the high temperature superconductors
to fulfill their promise as potentially useful materials.

%


\begin{acknowledgments}
This work was supported  in part by NSF (OISE-0352598) and by
DST (India).
\end{acknowledgments}


\end{document}